# Quantitative explanation of the reported values of the Hall-Petch parameter

Y. Li,[a] A.J. Bushby [b] and D.J. Dunstan [a *]

[a] School of Physics and Astronomy,
[b] School of Engineering and Materials Science,
Queen Mary University of London,
London, E1 4NS, England.

**Abstract:** The Hall-Petch effect has been described for the past sixty years as a dependence of the strength of polycrystalline metals on the inverse square-root of grain size $d$. The value of the coefficient of the dependence has been the subject of discussion throughout. Here, we find what factors in the experiments determine its values. A meta-analysis using maximum-likelihood methods is reported of the literature values of the coefficient in sixty-one datasets. Weak dependence is found on composition and bulk strength. Clear dependence is found on the elastic anisotropy of the different metals and on their stacking fault energies. Surprisingly, no dependence is found on plastic strain, and the strongest dependence is found on the average grain size of each study. Combining these effects accounts for the reported values of about 80% of the sixty-one coefficients. The grain-size dependence and the bulk strength dependence indicate that the Hall-Petch coefficient is an artefact arising from incorrect fitting of the data. Moreover, the grain-size dependence implies a minimum strength described by a simple inverse $1/d$ or a $\ln d/d$ function, which arises theoretically from considerations of dislocation curvature.

**Keywords:**
metals; grain boundary constraint; Hall-Petch effect; dislocations; scaling exponents.

\* **Corresponding Author**
Email address:      d.dunstan@qmul.ac.uk
Telephone           (+44) 020 7882 3411
Postal address:     School of Physics and Astronomy,
                    Queen Mary University of London,
                    London, E1 4NS, England.

# Graphical Abstract

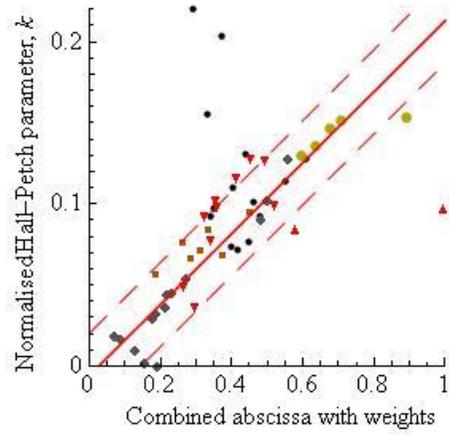

$$\sigma = \sigma_0 + \frac{k_{HP}}{\sqrt{d}}$$

The abscissa contains weighted factors:

$w_1$ stacking fault energy
$w_2$ composition
$w_3$ bulk strength
$w_4$ elastic anisotropy
$w_5$ grain size

and explains most of the reported $k_{HP}$ values

# 1. Introduction

The Hall-Petch equation expressing the inverse-square-root dependence of yield or flow strength of polycrystalline metals on the grain size was first proposed in 1951 by Hall [1] and confirmed in 1953 by Petch [2]. Since then, a large body of experimental data has been published in good agreement with this equation. Although it has frequently been challenged, as early as 1958 by Baldwin [3], it is presented in most elementary materials textbooks and university courses and used in industry to predict strength. Most authors consider that it has strong experimental support. Indeed, a conference was held in 2013 to celebrate its sixtieth anniversary [4], with a conference logo incorporating the Hall-Petch equation,

$$\sigma_Y = \sigma_0 + kd^{-1/2} \qquad (1)$$

In this expression, $\sigma_Y$ is the yield strength, but the equation has been used as often to describe the flow stress at a given plastic strain, $\sigma(\varepsilon_P)$, and in what follows we will not need to distinguish yield stress and flow stress [5, 6]. The constant $\sigma_0$ is the yield or flow stress of single-crystal or bulk large-grain-size polycrystalline material. It is expected to be largely dependent on the history and preparation of the specimen, apart from having, for each metal, a minimum value in ideal specimens due to the Peierls stress. The second term on the right-hand side of Eq.1 describes the dependence of yield or flow stress on grain size $d$. The notation $d_{ISR}$ will be convenient below for the inverse square-root of grain size, $d^{-\frac{1}{2}}$.

The values of the parameter $k$ for different metals have been of great interest throughout this period. Experimentally, values reported in the literature even for the same metal are often very scattered. Many authors have proposed theories consistent with the inverse square-root of $d$ in Eq.1 and capable of explaining the experimental values of $k$. See, e.g., the early review of Li and Chou [7], the comprehensive review of experiments and theories of Cordero *et al.* [8], and references therein, and two modern discussions specifically of the value of the parameter $k$ [9, 10]. Such attempts have been frustrated, either by the large scatter in the experimental values of $k$, or by lack of clarity as to the applicability of the predictions of the various theories to specific experimental situations. Thus, some theories predict a non-zero $k$ for the yield point. Others predict $k = 0$ for the yield point and $k$ proportional to plastic strain $\varepsilon_{pl}$ for the flow stress. It may be debated whether $k$ is a derived quantity predictable from more fundamental material parameters, or whether it is a material parameter in its own right to be measured for each metal but not predictable from more basic considerations. Or, like $\sigma_0$, it may be largely dependent on the history and preparation of the samples.

In two previous papers, we have challenged Eq.1 [5, 6]. Following Baldwin [3], we found that the quality of fit of Eq.1 to many datasets was as good with exponents $x$ from $x = \frac{1}{4}$ to $x = 1$ as it was with the $x = \frac{1}{2}$ of Eq.1. Using dummy data sets and fitting for the value of the exponent $x$, we found that the least-square residuals fitting procedure is biased due to the random errors in grain-size estimation, and returns an exponent on average about half the true value. We considered also the probability of the data being where it is under the different hypotheses of Eq.1 and alternative expressions. In the course of that work, we gathered sixty-one datasets from the literature that have been considered to support Eq.1. Here, we report a meta-analysis of the sixty-one values of $k$ from these datasets to see what can be deduced from them about the representation of the data by Eq.1 and the physical interpretation of $k$. Note that in this analysis we do not invoke any theories of the Hall-Petch effect. Our purpose here is only to find what factors in the experiments influence or determine the observed values of $k$.

Our conclusions are that there is a clear dependence of $k$ on anisotropy, and on stacking-fault energy, and that there is no evidence of a dependence of $k$ on plastic strain. We find weak dependence on composition (purity) and bulk strength. Surprisingly, since the dependence of $\sigma$ on $d_{ISR}$ in Eq.1 is already given explicitly, there is a strong dependence of $k$ on $d_{ISR}$. That is, the coefficient $k$ is not a constant as it should be, but as the function $k(d)$ it conceals within itself much of the true functional dependence of $\sigma$ on $d$. These outcomes of the meta-analysis of course have implications for theories of Eq.1, discussed in Section 4.

## 2. Meta-analysis of $k$ values

We apply meta-analysis to the sixty-one values of $k$. While commonplace in social sciences and medicine, meta-analysis is relatively unusual in materials science and metallurgy (but see Deville *et al.* [11]). The purpose of meta-analysis is to take multiple studies and by combining their results to obtain a greater statistical significance for a result, or, less often, to obtain a result that the original studies did not consider. Meta-analysis has dangers, which can introduce bias. This is well-documented in the medical literature. See the Appendix for a discussion of their relevance and their mitigation here.

2.1. Data sources and selection.

We assembled a body of data consisting of data-sets that were fitted with Eq.1 by their authors or subsequently. Citations, references and search engines led us to more data, as did helpful input from colleagues. We included more recent data when we found it, but our emphasis was on the early data that contributed towards the establishment of Eq.1. All the data-sets that we found, we use; that is, there has been no selection. The sixty-one that we use are distinguished from the unknown number in the literature only by the random accidents of the search processes. There is therefore no risk of selection bias (see Appendix).

Different authors gave more or less information on specimen characterisation, measurement techniques, and errors, but in any case such information was not used by the original authors to correct in any way the data values fitted with Eq.1 nor the parameter values obtained from the fits. Consequently, it would be inappropriate – it would risk bias – to use any such information here. We work with the raw data.

2.2. Meta-Analysis Factors

Comparing the values of $k$ across many studies requires considering several major factors which are expected from theory to affect values of $k$. Data from both tension experiments and indentation hardness testing are used. We divide the hardnesses by a factor of 2.8 for comparability with the tension data, and attribute the nominal value of plastic strain $\varepsilon_{pl}$ = 0.2 to these datasets. A number of different metals are used. All of the theories of Eq.1 predict that $k_{HP}$ will depend on the elastic moduli $c_{IJ}$ of the metal and on its Burger's vector $b$. We normalise stresses for different metals by dividing by the Young's modulus to give elastic strains. Similarly the grain sizes are normalised to the size of the crystal unit cell, by dividing the values given by the lattice constant of each metal. (This may be taken as a proxy for normalising to the Burgers vector, which would introduce uncertainties as to the appropriate projections of the vector onto relevant slip planes, etc.) For details of the normalisation see Ref.6 and the Supplementary Information. Following these normalisations, the value of $k$ is dimensionless.

This leaves five known factors in the experiments which may affect the data. Different metals have different elastic anisotropies and this should affect how polycrystalline specimens behave. Some datasets report yield stresses, which ideally would be at a plastic strain of $\varepsilon_{pl}$ = 0 but may be at the conventional $\varepsilon_{pl}$ = 0.002 or at a lower or upper yield point, while others report flow stresses at various plastic strains $\varepsilon_{pl}$ up to 0.3 – and theories differ in their predictions of the variation of $k_{HP}$ on plastic strain. Different datasets use widely varying ranges of grain size. The metals studied vary in their purity, or number of metallurgically significant elements, from commercial brass and steel to high-purity aluminium. The fitted bulk strength $\sigma_0$ may be treated as a factor. Any known physical properties of the various metals could also be considered as factors. We tested the stacking-fault energy, bringing the total of factors considered here to six.

In Ref.6, the data were first digitised, normalised as described above, and fitted with Eq.1. Full information on the datasets, normalisation and fitting are given in [6] and the data are given in the Supplementary Information. Here, to test the effects of these six factors, the values of $k$ returned by the fits are plotted against each factor in turn, and tested for correlation with each of the six factors by fitting the data to the function $y = ax + b$. Independence of $k$ of a given factor corresponds to fitted values of $a$ close to zero, within the statistical error bar.

2.3 Statistical Analysis

In a standard analysis of experimental results, data is obtained as a function of experimental parameters in the light of theory. There will be a predicted functional dependence and perhaps quantitative predictions of coefficients, and it is to test and refine these predictions that the experiments are performed. Meta-analysis proceeds differently. We have a set of reported data, here, values of $k$, and potential factor values, constituting a large matrix of numbers. The objective is to establish correlations within this matrix. The most powerful way to do this is factor analysis but for our purposes here it is preferable to use a less powerful but more transparent technique.

We begin by inspecting the properties of the 61 normalised values of $k$. They have a mean of 0.155, but a wide distribution of values from –0.001 to 0.998 (Fig.1), so the mean is near the lower end of the range. The standard deviation about the mean is 0.207.

The kurtosis (fourth moment over second moment) is 11 and the skewness is 3, compared with the values $3 \pm 0.5$ and $0 \pm 0.3$ expected for a Gaussian distribution of 61 numbers For $\log_{10}k$, the mean is $-1.02$, variance 0.214, kurtosis 7.5 and skewness $-1$. Thus the distribution of the data is far from normal or lognormal. Least-squares fitting methods assume that the residuals $r_i$ – the scatter of the data around the fitted model – are Gaussian-distributed, independently drawn from an identical normal distribution (i.i.d.). When that is not so, as here, least-squares methods discard much of the information in the data, and it is preferable to use other methods which make use of more of the information. We use Maximum Likelihood methods. For accessible introductions to these methods, see e.g. [12, 13].

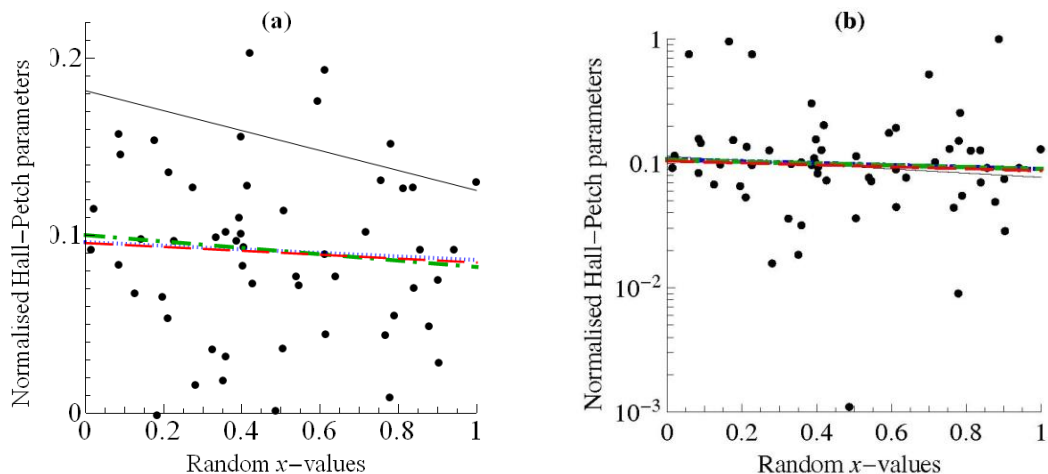

**Fig.1.** (Colour on-line.) The 61 values of the Hall-Petch parameter $k$ in (a) and $\log_{10}k$ in (b) are plotted against a random abscissa in the range 0-1. The extreme high values are not shown in (a) but may be seen in (b); similarly the extreme low and negative values are not shown in (b) but may be seen in (a). Fits are shown by a classic least-squares (LS) procedure (thin solid lines) and by Maximum Likelihood methods using the Lorentzian pdf (thick dotted blue lines, the double Gaussian pdf (DG, thick chain-dotted green lines) and the Gaussian plus flat pdf (GF, thick dashed red lines) with the values given in Table 1.

2.3.1. *Testing with dummy datasets.* We begin by testing the statistical methods against dummy datasets, as a benchmark. Suitable dummy datasets may be constructed as ($\varepsilon_i$, $k_i$) where $\varepsilon_i$ are random numbers in the range 0 to 1, and $k_i$ are the Hall-Petch parameters ($i = $ 1 to 61). Using a least-squares procedure to fit these dummy data to $ax + b$, we expect, within error, $a = 0$ and $b = \bar{k}$ where $\bar{k} = 0.155$ is the mean value of our collection of $k$ values. One such dataset gives $a = -0.056 \pm 0.098$, $b = 0.182 \pm 0.054$, as expected within error (Fig.1a). However, the correlation matrix for $a$ and $b$ has off-diagonal terms of about 0.9. It is better to fit to the datasets $(\varepsilon_i - \bar{\varepsilon}, k_i - \bar{k})$, and restore the average values after fitting. This gives the same result for $a$, with $b = \bar{k} - a\bar{\varepsilon} \pm 0.027$ and the off-diagonal terms in the correlation matrix are now zero. In this way, we created 500 such dummy datasets. The 500 values of *a* averaged to –0.002 and the standard deviation of the values of *a* was 0.055 (Table I). The value of *b* and the error in it is no longer interesting.

Given the presence of excess kurtosis, the few datapoints at values of *k* up to 1 are heavily distorting the least-squares fits. A common procedure in such cases is to eliminate these datapoints, as outliers, and fit to the remaining points which have a much narrower normal distribution. It is better to exploit all the information in the data using the Maximum Likelihood method, as any probability distribution function (pdf) may be used as best fits the data. For any pdf $P(x)$ for the residuals $r_i$, the likelihood of each datum $k_i$ is $P(r_i)$, and the likelihood of the dataset under the model is

$$L = \prod_{i=1}^{n} P(r_i), \quad \ln L = \sum_{i=1}^{n} \ln P(r_i) \qquad (2)$$

Residuals are functions of the parameters *a* and *b* of the model, and for a normal distribution of residuals centred on zero, the standard deviation $\sigma$ of the Gaussian pdf is the third parameter of the model. Values of $L$ can be very large or very small, and have no intrinsic meaning. It is more convenient to sum the natural logarithms of $P(r_i)$ to calculate the log-likelihood, $\ln L$. This is then maximised with respect to *a*, *b* and $\sigma$. Doing this with the single dummy dataset mentioned above, we get $a = –0.056$, as above, and $\sigma = 0.29$ which is not an error estimate for *a* or *b* but describes the variance ($\sigma^2$) of the data with respect to the model. The log-likelihood is $\ln L = 10$. Error estimates for *a* and *b* can be found by varying each individually while leaving the other as a free fitting parameter and looking for a reduction in $\ln L$ of 0.5. Here that gives $a \pm 0.096$ in agreement with the least-squares method as expected for a Gaussian pdf.

Changes in the fitted model that give increases in $\ln L$ of more than 2 or 3 are what matter, as this is equivalent to $4\sigma$ - $6\sigma$ in classic least-squares methods. Accordingly, we calculate $\ln L$ with other pdf's. Given the large kurtosis, we are interested in fat-tailed distributions compatible with a significant proportion of apparent outliers in the dataset. The Lorentzian is suitable. The maximum of $\ln L$ is found to be 56.5 for $a = – 0.018 \pm 0.024$, with width $\gamma = 0.034$. Thus *a* is found to be significantly closer to zero, by a factor of about three, than by the least-squares model (Fig.1a). The increase in $\ln L$ is massively significant – with 0.5 in $\ln L$ corresponding to $1\sigma$ in a Gaussian analysis, an increase in $\ln L \sim 2 – 3$ is considered to give very significant preference to one model over another, and here we have an increase of 45. Moreover, we do not have to consider that the datapoints that are a long way from the fitted line are outliers – on the contrary, they are accommodated by the Lorenztian pdf on a par with all the other datapoints.

Other fat-tailed pdfs give similar results. We can attribute the outliers to a broad Gaussian pdf and the other data points to a narrow one about the line $y = ax + b$, so that we have a double-Gaussian pdf, referred to below as DG. Then, as well as looking for the values *a*, *b* and the $\sigma$-values of the two distributions, we look for the fraction *f* of the broad pdf and the fraction $(1 – f)$ of the narrow pdf that maximizes $\ln L$. Fitting with the two extra parameters, the width of the second Gaussian and the value of the fraction *f*, we obtain $a = 0.010 \pm 0.025$, $f = 0.142 \pm 0.050$, and the standard deviations of the narrow and broad distributions are 0.067 and 0.80 respectively. That is, about one seventh of the data belong to the broad distribution, and the others form a distribution sharper by a factor of more than two than the single Gaussian PDF gave. The value of ln$L$ is 61. Each extra fitting parameter requires an increase in ln$L$ of 1 to offset it (this corresponds to the Akaike information criterion [14]), so this model is only slightly more probable than the Lorentzian. However, it provides a first insight into the number of apparent outliers – the fraction $f$ of the data, or about ten datapoints that appear to belong to a different pdf.

Alternatively, we can consider that each datum has a probability $f$ of being an outlier in this sense, and that the outliers have a flat distribution over a range of $k$ values of width $g$, hence a flat pdf of magnitude $fg^{-1}$ over the range. Then $(1-f)$ of a sharp Gaussian distribution (the GF pdf) is added. Now, maximization of ln$L$ yields $a = 0.011 \pm 0.025$, $f = 0.158 \pm 0.056$ and ln$L = 68.5$ (Fig.1a). Especially with one fewer parameter than the double Gaussian, this increase in ln$L$ is highly significant – this is by far the best model.

**Table 1.** Results of fitting randomized datasets with various Maximum Likelihood (ML) probability distributions, Gaussian (G), equivalent to a least-squares (LS) fit, Lorentzian (Lor), double-Gaussian (DG) and Gaussian plus flat (GF).

|  | ML/MS G | ML Lor | ML DG | ML GF |
|---|---|---|---|---|
| $P(r_i)$ | $\dfrac{1}{\sigma\sqrt{\pi}}e^{-r_i^2/\sigma^2}$ | $\dfrac{1}{\pi}\dfrac{\gamma}{r_i^2+\gamma^2}$ | $fG(\sigma_1)$ $+(1-f)G(\sigma_2)$ | $fg+(1-f)G(\sigma)$ |
| 500 LS dummies | $a$: –0.003 ± 0.092 | | | |
| Dummy D1 | ln$L$: 10.0<br>$f$: 0<br>$a$: –0.056 ± 0.096<br>$\sigma$: 0.29 | ln$L$: 56.6<br>$f$: 0<br>$a$: –0.018 ± 0.024<br>$\gamma$: 0.034 | ln$L$: 61.2<br>$f$: 0.14 ± 0.5<br>$a$: –0.010 ± 0.025<br>$\sigma_1$: 0.07; $\sigma_2$: 0.8 | ln$L$: 68.5; $g$: 1<br>$f$: 0.15 ± 0.05<br>$a$: –0.011 ± 0.025<br>$\sigma$: 0.062 |
| log$_{10}$D1 | ln$L$: –38.0<br>$f$: 0<br>$a$: –0.17 ± 0.21<br>$\sigma$: 0.65 | ln$L$: –29.2<br>$f$: 0<br>$a$: –0.072 ± 0.11<br>$\gamma$: 0.16 | ln$L$: –26.5<br>$f$: 0.42 ± 0.12<br>$a$: –0.076 ± 0.11<br>$\sigma_1$: 0.21; $\sigma_2$: 1.0 | ln$L$: –25.1<br>$f$: 0.09 ± 0.03; $g$: 3<br>$a$: –0.078 ± 0.12<br>$\sigma$: 0.27 |

Similar results are obtained by fitting to values of log$_{10}k_i$ (Fig.1b). The ordinary least-squares method works better here, as there is less skewness in this distribution, but still gives an error in the gradient nearly three times the gradient given by the GF pdf. The least squares ln$L$ is –39, rising to –29.5 for the Lorentzian, –26 for the double-Gaussian and –25 for the GF pdf. (The absolute value of ln$L$ is not important; it is lower here because the pdf is spread more thinly on the $y$-axis from –3 to 0 instead of 0 to 1, corresponding to $g = 3$ in Table 1). The main difference is the much higher proportion of outliers attributed to the log$_{10}k_i$ by the DG and GF pdfs. It probably occurs because the true outliers are much closer to the main distribution, as seen in the DG results by the relative values of $\sigma_1$ and $\sigma_2$.

What we have established in this Section is that the Maximum Likelihood methods are about four times more sensitive than the least-squares method for exposing a correlation or lack of correlation between the data, the experimental values of $k$, and the abscissa or factor against which they are plotted. The key benchmark is the variance of the fitted values of the slope, $a$. Using least-squares fitting, a slope $a$ of 0.096 (on a plot where the abscissa values have been normalized to the range 0–1) is not significant. Using Maximum Likelihood methods, this criterion is sharpened to about 0.024. Of these methods, the flat distribution for the outliers with a Gaussian for the bulk of the data is the most probable model. There is a consensus among the methods that about 15% of the data, or ten of them, are outliers. The GF pdf gives the highest likelihoods. The $k_i$ and the $\log_{10} k_i$ give very similar results, except that the outliers are probably over-reported in the $\log_{10} k_i$ analysis.

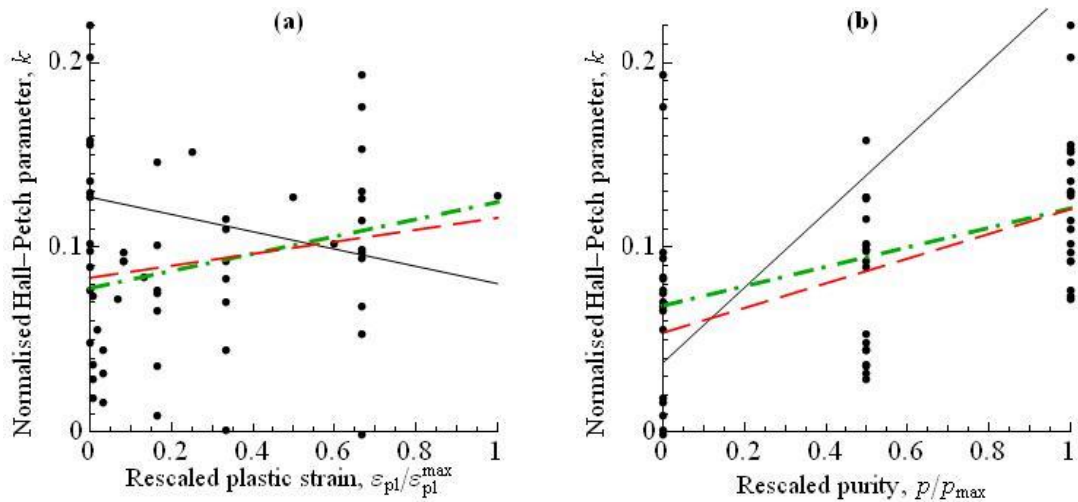

**Fig.2.** (Colour on-line.) Plots and fits against (a) plastic strain $\varepsilon_{pl}$ and (b) our purity factor $p$. The thin solid lines are the least-squares fits, the thick chain-dotted green lines are the fits using the Lorentzian pdf and the thick dashed red lines are the fits of the GF model.

2.3.2. *Physical Factors:* The next step is to plot the data against the various physical factors reported by the original authors and to fit with $ax + b$ as for the dummy datasets of the previous Section. For comparability, the range of each factor is rescaled to the dimensionless range 0-1. The resulting slope $a$ for each factor is given in Table 2 for the fits to each of the six factors plotted in Fig.2-Fig.4. The corresponding slopes $\alpha$ for fits to the real values of the factors are also given. It should be noted, however, that the slopes $\alpha$ are dimensionless (except for stacking-fault energy) because of our initial normalizations of yield or flow stress by division by the Young's modulus to yield or flow elastic strain, and grain size to number of units cells per grain by division by the lattice constant. This initial normalization was essential if different metals are to be compared.

Some readers will need to be warned that the large scatter in these plots does not show that they mean nothing. Such eye-balling is not a statistical technique mathematically proven to extract the full information in the data. The Maximum

Likelihood method is [12], and what matter is ln*L*, not a subjective assessment of the scatter.

To plot the data against plastic strain $\varepsilon_{pl}$, we deleted three datasets for brass, for which we could not find information on the strain. The fits to the remaining 59 values of *k* are shown in Fig.2a. While the least-squares method gives a negative slope – but zero within error – the maximum likelihood methods are very consistent in giving a positive slope outside error. At less than $2\sigma$, this might be interpreted as evidence for no effect of strain on the Hall-Petch parameter, however, the increase in log-likelihood to 72.6 (Table 2) from the random-variable value of 68.5 (Table 1) offers some support. The value of $\alpha_\varepsilon$ for this fit is meaningful, and dimensionless because *k* and strain are both dimensionless. To describe the purity or composition of the metals studied, given the complexities of metallurgy, we adopted the simple scheme of assigning the value $p = 0$ to a pure metal (four or five nines), $p = 1$ for the addition of an alloying element (as in brass) or purities around two nines, and $p = 2$ for anything more complicated, i.e. commercial iron and steel. In Fig.2b, there is a strong effect, with a slope at over $4\sigma$. The jump in the log-likelihood to 76 (Table 2) is highly significant. The value of $\alpha$ for this fit is not meaningful because of our arbitrary quantification of purity *p*.

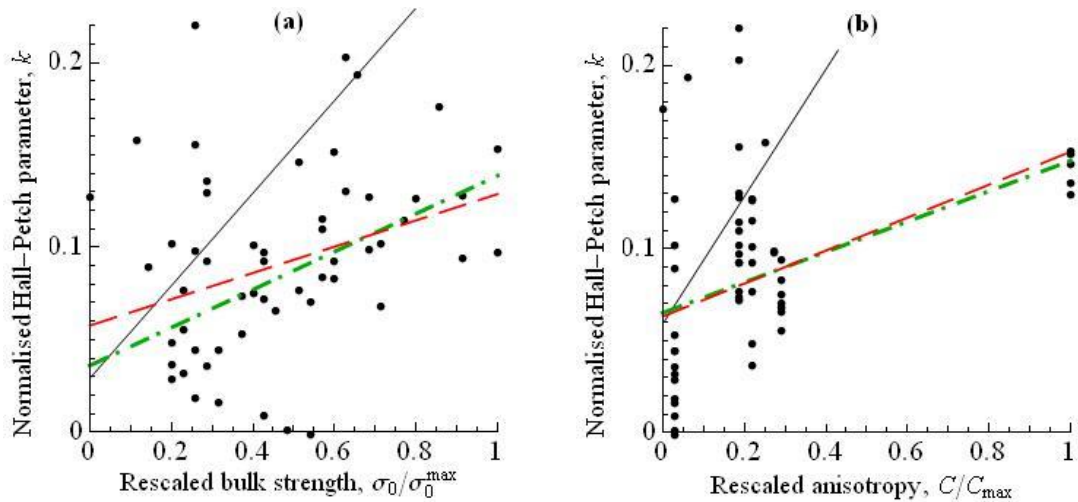

**Fig.3.** (Colour on-line.) Plots and fits against (a) bulk strength $\sigma_0$ and (b) anisotropy *C*. The thin solid lines are the least-squares fits, the thick chain-dotted green lines are the fits using the Lorentzian pdf and the thick dashed red lines are the fits of the GF model.

For the bulk strength (Fig.3a), we used the values of $\sigma_0$ obtained by fitting the data to Eq.1 (see Supplementary Information). Because $\sigma_0$ is normalized as elastic strain, the value of $\alpha$ for this fit, $\alpha_\sigma$, is dimensionless. For anisotropy, we deleted the titanium data, leaving 59 datasets, as the anisotropy factor is less well defined for hexagonal metals than it is for the cubic metals. For the cubic metals, we used the standard definition that is based on the ratio of the two shear moduli, $C = 2c_{44}/(c_{11} - c_{12}) - 1$. This has the advantage, compared with e.g. $C = 2c_{44} - c_{11} + c_{12}$, of being already normalised for different metals. It shows a stronger effect (Fig.3b), at almost $5\sigma$, but a rather surprising fallback in ln$L$ to 71 (Table 2). Because $C$ is dimensionless, the value of $\alpha$ for this fit, $\alpha_C$, is dimensionless.

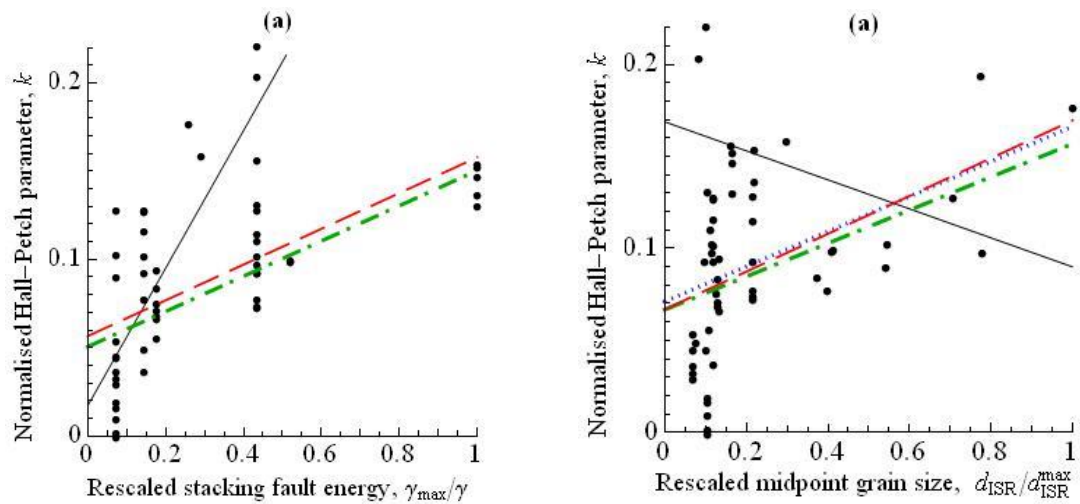

**Fig.4.** (Colour on-line.) Plots and fits against (a) the inverse of the stacking fault energy, and (b) the midpoint grain size of each dataset. The thin solid lines are the least-squares fits, the thick chain-dotted green lines are the fit using the Lorentzian pdf and the thick dashed red lines are the fits of the GF model.

Estimates of stacking-fault energies $\gamma$ were found in the literature for all the metals except Ti and Cr (see Supplementary Information). Deleting those three datasets left 58 values. Their slope against $\gamma$ is negative, so for easier comparability with the other factors, we used the inverse stacking-fault energy, $\gamma^{-1}$, plotted in Fig.4a. The slope $\alpha_{SFE}$ for this factor therefore has units of mJ m$^{-2}$. For the grain size, we take the mid-points of the datasets on the classic Hall-Petch plots, i.e. $d_{ISR}^{mean} = \frac{1}{2}(d_{ISR}^{max} + d_{ISR}^{min})$ (Fig.4b). Because the grain size is normalized to the number of lattice constants per grain, the value of $\alpha$ for this fit, $\alpha_d$, is dimensionless. This characterization of $d$ is motivated by consideration of what a $d^{-1}$ dependence looks like on a Hall-Petch plot. It is a parabola, and the slope of a straight line fitted to data that follow the parabola would be approximately the gradient of the parabola at the mid-point of the data range – though any dependence of $k$ on any measure of $d$ would contradict Eq.1. Yet this factor and the stacking-fault energy have the strongest effects, with the largest slopes, at 5$\sigma$, and with jumps in ln$L$ to 74 and 76 (Table 2).

Compared with the null hypothesis of Section 2.3.1, i.e. the hypothesis that there is no effect of these factors on the Hall-Petch parameter $k$ so that random ordering would give the same results as ordering the $k$ values by these factors, all six of these models are supported by the data. That being so, we should suppose they all represent effects that apply in all Hall-Petch experiments, so we should test the effect of these factors acting together. The ten or so outliers may be the result of other large, rare perturbations but the factors for these have not been identified.

**Table 2.** Fitting parameters for the various factors of Figs 2-4 using the Maximum Likelihood method with the Gaussian plus Flat pdf. The fraction of outliers is $f$ and the maximum log-likelihood is $\ln L$.

| Abscissa | Slopes $a$ | Slopes $\alpha$ | Outliers $f$ | $\ln L$ |
|---|---|---|---|---|
| Strain, $\varepsilon_{pl}$ | $a_\varepsilon = 0.033 \pm 0.020$ | $\alpha_\varepsilon = 0.1$ | $0.10 \pm 0.05$ | 73 |
| Purity, $p$ | $a_p = 0.067 \pm 0.015$ | $\alpha_p = 0.033$ | $0.16 \pm 0.06$ | 76 |
| Bulk strength, $\sigma_0$ | $a_\sigma = 0.073 \pm 0.028$ | $\alpha_\sigma = 20$ | $0.15 \pm 0.06$ | 72 |
| Anisotropy, $C$ | $a_C = 0.095 \pm 0.020$ | $\alpha_C = 0.012$ | $0.20 \pm 0.06$ | 71 |
| Stacking fault energy, $\gamma$ | $a_\gamma = 0.101 \pm 0.020$ | $\alpha_\gamma = 1.32$ m$^2$ mJ$^{-1}$ | $0.16 \pm 0.06$ | 76 |
| Grain size, $d_{ISR}$ | $a_d = 0.103 \pm 0.020$ | $\alpha_d = 3.9$ | $0.09 \pm 0.05$ | 74 |

It is worth considering how this would work in the ideal case. Suppose that these six factors are the only factors determining $k$, that they are wholly independent, and that their linear ($ax + b$ with $b = 0$) contributions to $k$ simply sum. Then in the fits to single factors so far discussed, the other five factors contribute both the intercepts $b$ and the scatter which gives the uncertainties in $a$. If we suitably scale all six factors and add their contributions to $k$, the intercept $b$ should decrease to zero and the uncertainties in $a$ should decrease; $a$ itself should increase. Keeping the abscissa normalized to the range 0 to 1, with all six factors included, the slope $a$ should be within error equal to the highest values of $k$ not belonging to outliers, i.e. about 0.18 or more if some of the outliers are brought within the main distribution when all factors are considered. The remaining outliers will be the result of other large, rare perturbations due to unidentified factors.
.
The anisotropy factor was rescaled to the 0-1 range by dividing values of $C$ by the highest value of $C$ in the data, 7.57, so the true gradient $\alpha_C$ is the value of $a_C$ reported above, divided by 7.57. The values of $d_{ISR}^{mean}$ were divided by 0.026 to rescale to the 0-1 range, so the true dependence is $k = \alpha_d\, d_{ISR}$ $k = \alpha_d d_{ISR}^{mean}$ with $\alpha_d$ given by $a_d / 0.026$. The gradients $a$ for strain, purity, $\sigma_0$ and stacking fault energy are similarly divided by 0.3, 2 and 20 and multiplied by 13 respectively to get $\alpha_\varepsilon$, $\alpha_p$ and $\alpha_\sigma$. The combined abscissa is $\alpha_C\, C + \alpha_d\, d_{ISR} + \alpha_\varepsilon\, \varepsilon_{pl} + \alpha_p\, p + \alpha_\sigma\, \sigma_0 + \alpha_\gamma\, \gamma^{-1}$. For comparison with the foregoing single-factor fits and plots, we rescale this new abscissa again to the 0-1 range. In order to include all sixty-one data, the missing values for plastic strain (brass), anisotropy (Ti) and the stacking fault energy (Cr) were allocated mid-range values (in italics in the Supplementary Information). The results of fitting to this dataset are given in Table 3. The slope $a$ is increased, and, most significantly, the $\ln L$ increases sharply to 88. This is a much better model than any of the six factors taken alone.

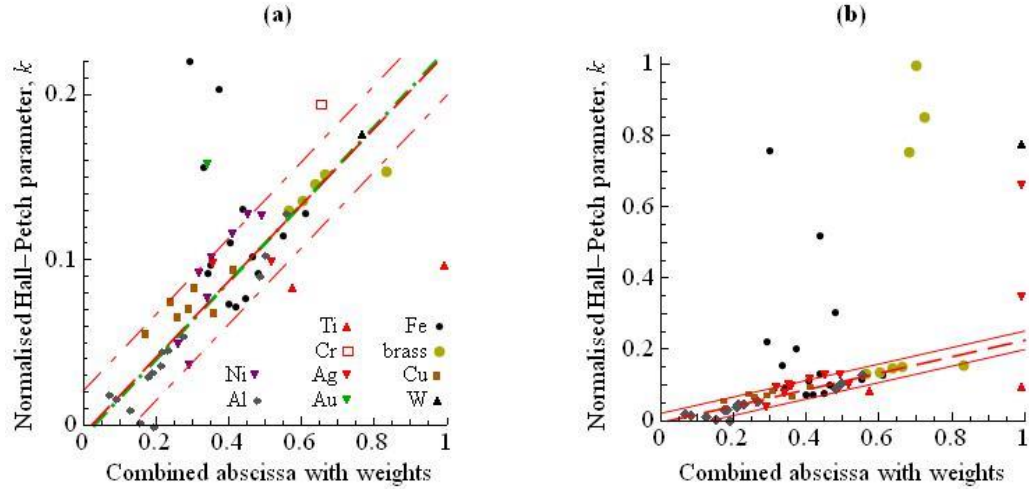

**Fig.5.** (Colour on-line.) Plots and fits against the weighted combination of all six factors of the last line of Table 3 (see text). In (a), the scale is the same as in Figs 2-4 for comparison; in (b) the scale is enlarged to show all the data. The thin solid lines are the least-squares fit, the thick chain-dotted green lines are the fits using the Lorentzian pd σf and the thick dashed red lines are the fits of the GF model. The thin red solid lines show the width (one sigma) of the Gaussian distribution of the mainstream data around the best GF fit.

**Table 3.** Fitting parameters for weighted combinations of factors. Fixed fitting parameters are in bold. In the first column, equal weights are given to all factors. In the final column, weights are optimized for the highest ln$L$. In between, each factor in turn has been eliminated by giving a fixed weight of zero and ln$L$ has been optimized with respect to the other weights. The final row gives the Akaike Information Criterion (AIC [15]) values for each model.

| Factors | $w_i$ | $w_i$ | $w_i$ | $w_i$ | $w_i$ | $w_i$ | $w_i$ | $w_i$ |
|---|---|---|---|---|---|---|---|---|
| Strain, $\varepsilon_{pl}$ | **1** | 0.2 | 1.2 | 1.3 | 2.2 | 2.6 | **0** | $0.7 \pm 0.8$ |
| Purity, $p$ | **1** | 2.2 | 0.8 | 0.9 | 1.1 | **0** | 1.5 | $1.3 \pm 0.3$ |
| Bulk strength, $\sigma_0$ | **1** | 1.8 | 0.7 | 0.4 | **0** | –0.4 | 1.5 | $1.1 \pm 0.7$ |
| Anisotropy, $C$ | **1** | 3.4 | 1.2 | **0** | 2.0 | 0.3 | 2.5 | $2.3 \pm 0.5$ |
| Stacking-fault energy, $\gamma$ | **1** | –1.5 | **0** | 1.2 | –1.0 | 1.6 | –1.6 | $-1.4 \pm 0.5$ |
| Grain size, $d_{ISR}$ | **1** | **0** | 2.0 | 2.2 | 1.8 | 1.9 | 2.2 | $2.0 \pm 0.2$ |
| "Outliers", $f$ | 0.19 | 0.24 | 0.20 | 0.16 | 0.20 | 0.16 | 0.21 | 0.20 |
| Slope, $a$ | 0.152 | 0.175 | 0.167 | 0.155 | 0.262 | 0.146 | 0.238 | 0.231 |
| Offset, $b$ | –0.06 | 0.022 | 0.006 | 0.019 | 0.0002 | 0.029 | –0.006 | –0.005 |
| Gaussian width, $\sigma$ | 0.037 | 0.030 | 0.029 | 0.038 | 0.027 | 0.039 | 0.026 | 0.026 |
| Log-Likelihood, ln$L$ | 88 | 87 | 96 | 93 | 98 | 90 | 99 | 100 |
| AIC | –158 | –158 | –176 | –170 | –180 | –164 | –182 | –182 |

However, we can add weightings $w_i \neq 1$ to the factors, while keeping $\Sigma w_i = 6$. The $\ln L$ is then maximized with respect to the six weights $w_i$ (five extra free fitting parameters) as well as to $a$, $b$ and $\sigma$. The optimum combined abscissa in Fig.5 uses the weightings of the last column of Table 3. This gives another large increase in the log-likelihood to 100, an increase in the slope $a$ to 0.231, and a very small intercept of $b = 0.006 \pm 0.0060$. The errors on the weightings are obtained as above, by fitting with each weight fixed in turn at its optimized value plus or minus an offset, and looking for the value of the offset which reduces $\ln L$ for the fit by 0.5. The scatter in the weightings across the Table is the consequence of eliminating each factor in turn by settings its weight to zero and then optimizing the other five.

We may expect reduced weightings of factors which are interdependent or correlated. For example, strain causes strain-hardening which increases strength. Equally, correlations with no physical significance will arise from experimenters' deliberate or chance choices – for example, it is possible that only metals with small stacking-fault energies were studied at high strain. If two factors are interdependent or correlated, their slopes in Figs 2-4 would each have a contribution from the other, and the weights in Table 3 would be reduced accordingly.

The last row of Table 3 gives the Akaike Information Criteria (AIC) values for each model. For choosing between models with different numbers of parameters, this quantity corrects the log-likelihood for the number of free fitting parameters in different models. It is defined as $2P - 2\ln L$ where $P$ is the number of parameters [15]. Its absolute values have no significance, but a decrease of $n$ of one model compared with another is equivalent to an $n$-sigma preference for the model with the lower AIC. Thus the models of the last two columns are to be preferred over the others, but there is nothing to choose between them. The weights in these two columns are all consistent within error, and so there is no evidence for a non-zero weight for plastic strain.

### 3. Discussion

Several points stand out among the data in Table 3. First, the weightings of plastic strain depend enormously on what other factors are taken into account. It is clear that, as a factor in determining the Hall-Petch coefficient, it is highly correlated with other factors. Eliminating plastic strain as a factor (penultimate column) reduces the $\ln L$ very little, and the AIC not at all. The models of the last two columns are highly preferred over all the others by their AICs, and whether plastic strain is deleted as a relevant factor or given a weighting consistent with zero within the error bar as in the final column does not matter. So, despite its importance in some theories of the Hall-Petch effect, we conclude that plastic strain is not a significant factor in determining the Hall-Petch coefficient.

The weightings of the four factors purity, bulk strength, anisotropy and stacking-fault energy swing violently according to which of them are included in or deleted from the analysis. These factors are closely related, that is, there are strong correlations among them. The bulk strength is clearly the least meaningful. Deleting it affects the AIC rather little, and in the last column its weight has an error nearly as large as the value. However, the methods here are not powerful enough to determine from this cluster of factors what are the real factors. Anisotropy, however, is clearly not ruled out, with its relatively large weighting and small error in the final column.

The weights for grain size in Table 3 are large and very stable against the selection of other factors. In the final column, it has the smallest error on its weight of all the factors. This indicates that it is largely independent of, and uncorrelated with, the other factors. Its large slope and large weighting combine to make it the most significant factor in determining the Hall-Petch coefficient.

The models with the highest ln$L$ are the combined fits with variable weightings of all six parameters of the last column of Table 3, and with plastic strain deleted as a factor (penultimate column). From these, the dominant factor in the Hall-Petch parameter $k$ is the grain size, $d_{ISR}$. (This is the midpoint grain size of the dataset, not of each specimen: the measure of grain size that matters for each specimen must be its own $d_{ISR}$; it is not possible for the strength of a single specimen to depend on the range of grain sizes that determined $d_{ISR}^{mean}$.) This is in direct contradiction of Eq.1, in which the effect of $d_{ISR}$ on $\sigma$ is supposed to be entirely in the denominator. That the bulk yield or flow stress, $\sigma_0$, should play a role is also in direct contradiction of Eq.1 in which it is only added to the Hall-Petch effect. Considering only the factor $d_{ISR}$, we have,

$$\sigma_Y = \sigma_0 + \frac{k}{\sqrt{d}} = \sigma_0 + \frac{\alpha_d d_{ISR}}{\sqrt{d}} = \sigma_0 + \frac{\alpha_d}{d} \qquad (3)$$

To see the effect of the strength, $\sigma_0$, consider a family of curves $\sigma_Y = \sigma_0 + a_d/d$ on a log-log plot with $a_d$ constant and $\sigma_0$ varied). The position of the elbow between the part of the curve with a slope near $-1$ at small $d$ and the part with a slope near 0 at large $d$ (i.e. around $d \sim a_d/\sigma_0$) moves to smaller $d$ as $\sigma_0$ is increased, at one decade per decade. A similar family of curves $\sigma_Y = \sigma_0 + k\, d_{ISR}$ has the corresponding elbows move to smaller $d$ at two decades per decade. Consequently, if data obeying the former are fitted with the latter family of curves, to keep the elbow in the right place to fit the data, it will be necessary to increase $k$ at one decade per decade of $\sigma_0$. That is, the coefficient will increase linearly with $\sigma_0$, as observed here. So both of these effects, the dependence of $k$ on $d_{ISR}$ and on $\sigma_0$, occur naturally if the true dependence of yield or flow stress, $\sigma_Y$, on grain size goes as $1/d$ (or ln $d$ / $d$) yet is fitted by a linear dependence on $d_{ISR}$.

As well as the six factors considered here which do contribute to the Hall-Petch parameter $k$, there are phenomena such as dislocation pile-up which are thought to contribute to the Hall-Petch parameter [15, 16]. The analysis here gives little room for these other parameters – only the value of the intercept, $b = 0.006 \pm 0.006$, gives any scope for them The theory of pile-up does not predict any dependence on the factors considered here, but (in normalized units) is supposed to contribute a value $p$ up to about 0.03 to $k$, a value which is capped by the theoretical strength of materials [16]. What is clear, is that even if pile-up were to account for the intercept, it has little to do with the values of $k$ ranging from 0.01 up to 0.2.

It might be thought that these results show that the Hall-Petch equation should be rewritten as

$$\sigma_Y = \sigma_0 + kd_{ISR} = \sigma_0 + \frac{0a_\varepsilon + 1.5\alpha_p + 1.5\alpha_\sigma + 2.5\alpha_C - 1.6\alpha_\gamma}{6}\frac{1}{\sqrt{d}} + \frac{2.2\alpha_d}{6}\frac{1}{d} \quad (4)$$

(α values from the penultimate column of Table 3). That would be incautious. The path from the raw data to Eq.4 has gone *via* an invalid fitting procedure (fitting to Eq.1), generating an invalid constant *k*, followed by a correction in the light of Fig.4b. There is no evidence in the results reported here that the correction will not have wider effects; no evidence that the middle term, over the square-root of *d*, in the RHS of Eq.4 is correctly formulated. It will be preferable to start afresh not from data (*k*) resulting from fitting to the incorrect Eq.1, but from the raw data, that is, the measured values of σ and the values of all known factors, and to carry out rigorous factor analysis. We have not done that here, because that would merely set up a rival fit to compete with Eq.1. The technique we use here is less powerful, but is better targeted to reveal unambiguously and transparently the issues with Eq.1. It is unambiguously shown to be the incorrect equation to fit the data, by the resulting internal contradictions revealed by meta-analysis.

There remains a factor (or factors) that we have not identified, that would account for the 10% or 15% of reported *k* values, up to *k* ~ 1, that are many standard deviations away from Eq.4 (Fig.5b). The key point here is that we kept these data in the analysis, so that if any of the factors had accounted for them (or if the Lorentzian pdf had been the pdf with the highest likelihood), they would have rejoined the main pdf rather than keeping their own separate pdf.

## 4. Conclusions

The variation of reported (normalised) values of the Hall-Petch coefficient *k* that fall in the range 0 to 0.2 is largely accounted for by the various factors reported by the original authors – specifically, by the purity, the bulk strength and the grain size, and also by the material parameters, the elastic anisotropy and the stacking-fault energy. The very wide variation of the remaining 20% of reported values in the range up to 1 remains unaccounted for. The variation in values of *k* that arises from the various values of the different factors in different studies does not of itself indicate that Eq.1 is incorrect. However, that the grain size itself is one of the factors that matters most, unambiguously demonstrates that fitting the data to Eq.1 is incorrect. Rather remarkably, plastic strain is not a factor.

Some comment can be made on the consequences of these conclusions for the various theories of the Hall-Petch effect. It is now clear that the difficulties previously encountered in explaining the values of the Hall-Petch coefficient *k* are due to trying to explain an incorrect equation. The theories that invoke plastic strain will need to be reconsidered, perhaps in terms of the cluster of correlated factors more-or-less related to strength. The theories that do not invoke the factors found here, such as pile-up [1, 2, 15], are very restricted in their application, since they would account for an intercept on the ordinate of Fig.5, which here is zero within error. Grain size is the dominant independent factor, and the resulting dependence $\sigma_Y$ on $d^{-1}$ supports the theory of dislocation curvature or source size constraint [6].

The other factors can all go to zero in suitable circumstances. That leaves the grain-size itself as the only universal contributor to *k* and the general $1/d$ or $\ln d / d$ size effect as the only ubiquitous Hall-Petch mechanism, to which will be added the effects of the mechanisms that invoke the other factors when applicable.

**Acknowledgements**

Yuan Li is grateful to the Chinese Scholarship Council for his research studentship. This research did not receive any specific grant from funding agencies in the public, commercial, and not-for-profit sectors.

**Appendix.** Pitfalls of Meta-Analysis

The purpose of meta-analysis is to take multiple studies and by combining their results to obtain a greater statistical significance for a result, or, as here, to obtain a result that the original studies did not consider. A review by Walker *et al.* [17] identifies four critical issues for meta-analysis. They are discussed in more detail by Cooper *et al.* [18].

**The file-drawer problem,** or publication effect [19] is liable to occur when the meta-analysis is conducted to test the same hypothesis that the original authors were studying. Only, or predominantly, studies with positive outcomes are published, while all the studies with null outcomes languish unpublished in the filing-cabinet. Then, all the meta-study achieves is to confirm the original prejudice according to which positive results were interesting and null results not. That is not a risk here. The original authors did not select for publication only those datasets which fitted well with Eq.3. On the contrary, they might have rejected those that didn't fit Eq.1, but that would not matter to us. What matters is that they did not – could not – select data for publication according the fit with the hypothesis that we are testing.

The comprehensiveness of the search for studies also matters for if it is not comprehensive there is scope for selection bias in the studies selected for inclusion in the meta-study. Walker *et al.* [17] do not emphasise, though, that if the search not comprehensive, but is random with respect to the hypothesis under test, having fewer studies merely lowers the statistical significance of the result of the meta-analysis but does not invalidate it in any other way. That is clearly the case here, for the same reason as this study does not risk publication bias.

**Heterogeneity of results**, or not comparing like with like, risks burying a few positive results from well-focused studies under scattered results from many less irrelevant studies. That is not an issue here. All the datasets we used gave good fits to Eq.1.

**Availability of relevant information** is the third key issue, and it does apply here. Among these datasets are of course wide variations of techniques such as grain-size measurement and characterisation of texture. Such variations were not generally fully reported by the original authors and the data were not then and cannot now be corrected in any way for them. However, such variations did not affect the validity of the datasets as published or later used as support for Eq.1, and no more do they affect their validity under meta-analysis as evidence of a behaviour which refutes Eq.1. Indeed, selection or correction of the raw data is dangerous in meta-analysis because of the risk of introducing bias. Some authors gave information about, for example, measurement of grain size, while others did not. This is not a risk factor providing that (1) there is little likelihood that the reported information or lack of it is correlated any of the factors, and (2) that no attempt is made to correct some data in the light of this information while other data cannot be corrected for lack of the information.

**Analysis of data** is a rather technical issue that does not concern us here, for it covers issues such as data-mining, in which a large body of data, tested for a very large number of correlations, will by chance give some false-positive outcomes among the very large number of true-negatives. Here we are looking for and finding specific outcomes predicted by theory.

**Good physics:** The previous point raises a final criticism of meta-analysis, made by a referee of a previous version of this paper: good physics does not arise out of statistical analyses of large datasets. Here, of course, the physics does not arise out of the statistical analysis. The physics arises out of the Orowan-Bragg ideas about dislocation curvature, or out of the Eshelby-Frank-Nabarro pile-up ideas. Statistics are used merely to choose between them, in as rigorous a way as possible. It is for this reason that we do not use a coefficient of determination such as $R^2$ to test Eq.1 or Eq.3, but consider, simply, the question, what model is most likely given the data?